\documentstyle[prl,aps,floats,epsf]{revtex}

\begin{document}

\baselineskip=12pt

\twocolumn[\hsize\textwidth\columnwidth\hsize\csname@twocolumnfalse\endcsname

\title
{Level Crossing Analysis of Growing surfaces }

\author
{F. Shahbazi, $^{a}$, S. Sobhanian,$^{b,e}$, M. Reza Rahimi Tabar
$^{c,d,e}$, S. Khorram,$^{f}$, \\ G.R. Frootan,$^{b}$ and H.
Zahed,$^{b}$ }
\address
{\it $^a$ Dept. of Physics , Sharif University of Technology,
P.O.Box 11365-9161, Tehran, Iran.\\
$^b$ Dep. of Theoretical Physics and Astrophysics, Tabriz University, Tabriz 51664, Iran  \\
$^c$ CNRS UMR 6529, Observatoire de la C$\hat o$te
d'Azur,
BP 4229, 06304 Nice Cedex 4, France,\\
$^d$ Dept. of Physics, IUST, P.O.Box 16844, Tehran, Iran. \\
$^e$ Research Institute for Fundamental Sciences, Tabriz 51664,
Iran.\\ $^f$ Center for Applied and Astronomical Research, Physics
Departmant, \\Tabriz Univrsity-51664, Tabriz Iran.
\\
}

\maketitle

\begin{abstract}

We investigate the average frequency of positive slope $\nu_{\alpha}^{+}$, 
crossing the height $\alpha = h- \bar h$ in the surface growing processes. 
The exact level crossing analysis of the random deposition model
and the Kardar-Parisi-Zhang equation in the strong coupling
limit before creation of singularities are given. 
\end{abstract}
\hspace{.3in}
\newpage
]

\section{Introduction}

Due to the technical importance and fundamental interest, a great deal of
effort has been devoted to understanding the mechanism of thin-film growth
and the kinetic roughening of growing surfaces in various growth techniques.
Analytical and numerical treatments of simple growth models suggest, quite
generally, the height fluctuations have a self-similar character and their
average correlations exhibit a dynamic scaling form [1-6]. It is known that
to derive the quantitative information of the surface morphology one may
consider a sample of size $L$ and define the mean height of growing film $
\bar{h}$ and its roughness $w$ by [1]: 
\begin{equation}
\bar{h} (L,t) = \frac{1}{L} \int_{-L/2} ^{L/2} dx h(x,t)
\end{equation}
and 
\begin{equation}
w(L,t) =( \langle ( h - \bar{h})^2 \rangle )^{1/2},
\end{equation}
where $\langle \cdots \rangle$ denotes an averaging over different
realization (samples). Starting from a flat interface (one of the possible
initial conditions), it was conjectured by Family and Vicsek [7] that a
scaling of space by factor $b$ and of time by a factor $b^z$ ($z$ is the
dynamical scaling exponent), re-scales the roughness $w$ by factor $b^{\chi}$
as follows: $w( bL, b^zt) = b^{\chi} w( L,t)$, which implies that 
\begin{equation}
w(L,t) = L^{\chi} f(t/L^z).
\end{equation}
If for large $t$ and fixed $L$ $(t / L^z \rightarrow \infty)$, $w$ saturate
then $f(x) \rightarrow const.$, as $x \rightarrow \infty$. However, for
fixed large $L$ and $1<< t << L^z$, one expects that correlations of the
height fluctuations are set up only within a distance $t^{1/z}$ and thus
must be independent of $L$. This implies that for $x << 1$, $f(x)\sim
x^{\beta}$ with $\beta=\chi / z$. Thus dynamic scaling postulates that, $
w(L,t) \sim t^{\beta}$ for $1 << t << L^z $ and $\sim L^{\chi}$ for $t >> L^z
$. The roughness exponent $\chi$ and the dynamic exponent $z$ characterize
the self-affine geometry of the surface and its dynamics, respectively.

Here we introduce the level crossing analysis in the context of surface
growth processes. In the level crossing analysis we are interested in
determining the average frequency ( in spatial dimension ) of observing of the
definite value for height function $h-\bar{h}=\alpha $ in growing thin films, 
$\nu _\alpha ^{+}$, from which one can find the averaged number of crossing
the given height in sample with size L. The average number of visiting the
height $h-\bar{h}=\alpha $ with positive slop will be $N_\alpha ^{+}=\nu
_\alpha ^{+}L$. It can be shown that the $\nu _\alpha ^{+}$ can be written
in terms of joint probability distribution function (PDF) of $h-\bar{h}$ and
its gradient. Therefore the quantity $\nu _\alpha ^{+}$ carry the whole
information of surface which lies in joint PDF of height and its gradient
fluctuations. This work aims to study the frequency of positive slope crossing
(i.e. $\nu _\alpha ^{+}$) in time $t$ on the growing surface in a sample with size L.
 We introduce a quantity $N_{tot}^{+}$ which is defined as $
N_{tot}^{+}=\int_{-\infty }^{+\infty }\nu _\alpha ^{+}d\alpha $ to measure
the total number of crossing the surface with positive slop.
 The $N_{tot}^{+}
$  and the path which is constructed by growing
surface are in the same order. It is expected that in the stationary state the $N_{tot}^{+}$
to become size dependent. We determine the time and
height dependence of $\nu _\alpha ^{+}$ for two exactly solvable models,
random deposition model (RD) and Kardar-Parisi-Zhang (KPZ) equation in the
strong coupling limit and before creation of singularities (sharp valleys)
with short range forcing. It is shown that the RD-model and KPZ equation
have different $\nu _\alpha ^{+}$ but $N_{tot}^{+}$ scales with time as $
t^{1/2}$ in both models.

 In section 2 we discuss the connection
between $\nu _\alpha ^+$ and  underlying probability distribution functions
(PDF) of growing  surfaces. Exact expression of $\nu _\alpha ^+$ for the RD 
model and with short-range forcing is given in section 3.  In section 4 we
derive the integral representation of $\nu _\alpha  ^+$ for the KPZ equation
in 1+1 dimension and in the  strong coupling limit before the creation of
singularities. We summarize the results in section 5.

\section{ The Level Crossing Analysis of Growing Surface}

Consider a sample function of an ensemble of functions which make up the
homogeneous random process $h(x,t)$. Let $n_{\alpha}^{+}$ denote the number
of positive slope crossing of $h(x)- \bar h = \alpha$ in time $t$ for a
typical sample size $L$ (see figure $1$ ) and let the mean value for all the
samples be $N_{\alpha}^{+}(L)$ where 
\begin{equation}
N_{\alpha}^{+}(L)=E[n_{\alpha}^{+}(L)].
\end{equation}

Since the process is homogeneous, if we take a second interval of $L$
immediately following the first we shall obtain the same result, and for the
two intervals together we shall therefore obtain 
\begin{equation}
N_{\alpha}^{+}(2L)=2N_{\alpha}^{+}(L),
\end{equation}
from which it follows that, for a homogeneous process, the average number of
crossing is proportional to the space interval $L$. Hence 
\begin{equation}
N_{\alpha}^{+}(L)\propto L,
\end{equation}
or 
\begin{equation}
N_{\alpha}^{+}(L)=\nu^{+}_{\alpha} L.
\end{equation}
which $\nu_{\alpha}^{+}$ is the average frequency of positive slope crossing
of the level $h - \bar h =\alpha$. We now consider how the frequency
parameter $\nu_{\alpha}^{+}$ can be deduced from the underlying probability
distributions for $h - \bar h$. Consider a small length $dl$ of a typical
sample function. Since we are assuming that the process $h-\bar h$ is a
smooth function of $x$, with no sudden ups and downs, if $dl$ is small
enough, the sample can only cross $h - \bar h=\alpha$ with positive slope if 
$h-\bar h < \alpha$ at the beginning of the interval location $x$.
Furthermore there is a minimum slope at position $x$ if the level $h- \bar h
= \alpha$ is to be crossed in interval $dl$ depending on the value of $h-
\bar h$ at location $x$. So there will be a positive crossing of $h-\bar h
=\alpha$ in the next space interval $dl$ if, at position $x$,

\begin{equation}
h- \bar h < \alpha \hspace{.6cm} and \hspace {.6cm} \frac{d(h-\bar h)}{dl}>
\frac{\alpha-(h - \bar h) }{dl}.
\end{equation}

\begin{figure}
\epsfxsize=7.9truecm \epsfbox{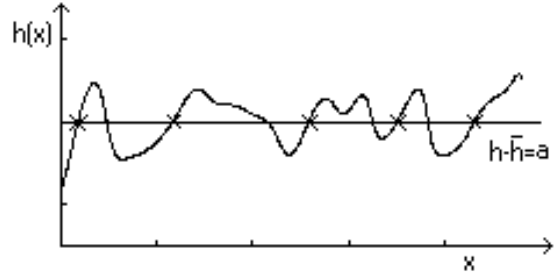} \narrowtext \caption{
positive slope crossing of the level $ h - \bar h = \alpha$.}
\end{figure}

Actually what we really mean is that there will be high probability of a
crossing in interval $dl$ if these conditions are satisfied [8,9].

In order to determine whether the above conditions are satisfied at any
arbitrary location $x$, we must find how the values of $y= h - \bar h $ and $
y ^{\prime}= \frac{ dy }{dl}$ are distributed by considering their joint
probability density $p(y,{y}^{\prime})$. Suppose that the level $y=\alpha$
and interval $dl$ are specified. Then we are only interested in values of $y
< \alpha$ and values of ${y}^{\prime}=(\frac{dy}{dl}) > \frac{\alpha-y}{dl}$
, which means that the region between the lines $y=\alpha$ and ${y}^{\prime}=
\frac{\alpha-y}{dl}$ in the plane ($y,{y}^{\prime}$). Hence the probability
of positive slope crossing of $y=\alpha$ in $dl$ is: 
\begin{equation}
\int_{0}^{\infty} d{y}^{\prime}\int_{\alpha-{y}^{\prime}dl}^{\alpha} dy p(y,{
y}^{\prime}).
\end{equation}

When $dl\rightarrow 0$, it is legitimate to put

\begin{equation}
p(y,{y}^{\prime})=p(y=\alpha,{y}^{\prime})
\end{equation}
Since at large values of $y$ and ${y}^{\prime}$ the probability density
function approaches zero fast enough, therefore eq.(6) may be written as:

\begin{equation}
\int_{0}^{\infty} d{y}^{\prime}\int_{\alpha-{y}^{\prime}dl}^{\alpha} dy
p(y=\alpha,{y}^{\prime})
\end{equation}
in which the integrand is no longer a function of $y$ so that the first
integral is just: $\int_{\alpha-{y}^{\prime}dl}^{\alpha} dy p(y=\alpha,{y}
^{\prime})=p(y=\alpha,{y}^{\prime}){y}^{\prime}dl $,  so that the
probability of slope crossing of $y=\alpha$ in $dl$ is equal to:

\begin{equation}
dl\int_{0}^{\infty} p(\alpha,{y}^{\prime}){y}^{\prime}d{y}^{\prime}
\end{equation}
in which the term $p(\alpha,{y}^{\prime})$ is the joint probability density $
p(y,{y}^{\prime})$ evaluated at $y=\alpha$.

We have said that the average number of positive slope crossing in scale $L$
is $\nu_{\alpha}^{+} L$, according to (7). The average number of crossing in
interval $dl$ if therefore $\nu^{+}_{\alpha}dl$. So average number of
positive crossings of $y=\alpha$ in interval $dl$ is equal to the
probability of positive crossing of $y=a$ in $dl$, which is only true
because $dl$ is small and the process $y(x)$ is smooth so that there cannot
be more than one crossing of $y=\alpha$ in space interval $dl$, Therefore we
have $\nu_{\alpha}^{+}dl=dl\int_{0}^{\infty}p(\alpha,{y}^{\prime}){y}
^{\prime}d{y}^{\prime}$, from which we get the following result for the
frequency parameter $\nu_{\alpha}^{+}$ in terms of the joint probability
density function $p(y , {y}^{\prime})$ 
\begin{equation}
\nu_{\alpha}^{+}=\int_{0}^{\infty}p(\alpha,{y}^{\prime}){y}^{\prime}d{y}
^{\prime}.
\end{equation}

In the following sections we are going to derive the $\nu_{\alpha}^{+}$ via
the joint PDF of $h-\bar h$ and height gradient. To derive the joint PDF we
use the master equation method [10-11]. This method enables us find the $
\nu_{\alpha}^{+}$ in terms of generating function $Z(\lambda,\mu,x,t) =
\langle \exp(-i\lambda (h(x,t)-\bar h)-i\mu u(x,t))\rangle$, where $u(x,t) =
- \nabla h$.

\section{ The Frequency of a Definite Height With Positive Slope for The
Random Deposition Model}

In random deposition model particles are dropped randomly over deposition
sites, and stick to the top of the pre-existing column on the site [ 1 ].
The height of each column thus performs an independent random walk. This
model leads to unrealistically rough surface whose overall width growth with
the exponent $\beta =\frac {1}{2}$ without saturation. In the continuum limit
the random deposition model is described by the following equations; 
\begin{equation}
\frac \partial {\partial t}h(x,t)=f(x,t)\hskip 1cm\frac \partial {\partial t}
u(x,t)=f_x
\end{equation}
where $h(x,t)$ is the height field, $u(x,t)=\frac \partial {\partial x}h(x,t)
$ and $f(x,t)$ is a zero mean random force gaussian correlated in space and
white in time, 
\begin{equation}
\langle f(x,t)f(x^{\prime },t^{\prime })\rangle =2D_0D(x-x^{\prime })\delta
(t-t^{\prime })
\end{equation}
where $D(x)$ is space correlation function and is an even function of its
argument. It has the following form; 
\begin{equation}
D(x-x^{\prime })=\frac 1{\sqrt{\pi }\sigma }\exp (-\frac{(x-x^{\prime })^2}{
\sigma ^2})
\end{equation}
where $\sigma $ is the standard deviation of $D(x-x^{\prime })$. The average force on
the interface is not essential and can be removed by a simple shift of $h$ (
i.e $h\rightarrow (h-\bar{F}t)$ where $\bar{F}=<f(x,t)>$) . Typically the
correlation of forcing is considered as delta function for mimicking the
short range correlation. We regularize the delta function correlation by a
gaussian function. When the variance $\sigma $ is much less than the system
size we would expect that the model would represent a short range character
for the forcing. So we would stress that our calculations are done for
finite $\sigma \ll L$, where $L$ is the system size. The parameters $D_0$ is
describing the noise strength.

Now assuming the homogeneity we define the generating function as: 
\begin{equation}
Z(\lambda ,\mu ,t)=\langle \exp (-i\lambda h(x,t)-i\mu u(x,t)\rangle 
\end{equation}
by using the eq.(14) we can find the following equation for the evolution of 
$Z(\lambda ,\mu ,t)$:

\begin{eqnarray}
\frac{\partial}{\partial t}Z(\lambda,\mu,t)&=&-i\lambda \langle
h_{t}(x,t)\exp(-i\lambda h(x,t)-i\mu u(x,t)\rangle \cr  \nonumber \\
&&-i\mu\langle u_{t}(x,t)\exp(-i\lambda h(x,t)-i\mu u(x,t)\rangle \cr 
\nonumber \\
&=&-i\lambda \langle f(x,t)\exp(-i\lambda h(x,t)-i\mu f_{x}(x,t)\rangle \cr 
\nonumber \\
&& -i\mu\langle u_{t}\exp(-i\lambda h(x,t)-i\mu u(x,t)\rangle \cr  \nonumber
\\
&=&-\lambda^{2}D_0 D(0)Z+\mu^{2} D_0 D_{xx}(0)Z.
\end{eqnarray}

The joint probability density function of $h$ and $u$ can be obtain by
fourier transform of the generating function:

\begin{equation}
P(h,u,t)=\frac{1}{2\pi}\int d\lambda d\mu e^{i\lambda h+i\mu
u}Z(\lambda,\mu,t),
\end{equation}

so by fourier transforming of the eq.(18) we get the Fokker-Planck equation
as:

\begin{equation}
\frac{\partial}{\partial t}P= D_0 D(0)\frac{\partial ^2}{\partial h^2}P-D_0
D_{xx}(0)\frac{\partial^2}{\partial u^2}P.
\end{equation}
the solution of the above equation can be separated as $
P(h,u,t)=p_{1}(h,t)p_{2}(u,t)$. Using the initial conditions $
P_{1}(h,0)=\delta(h)$ and $P_{2}(u,0)=\delta(u)$ (starting from flat
surface) it can be shown that:

\begin{eqnarray}
P(h,u,t) &=& \frac {1}{4\pi t\sqrt{-D_0^2D(0)D_{xx}(0)}}\cr \nonumber  \\
&&\exp (-\frac{h^2}{4D_0D(0)t}+\frac{u^2}{4D_0D_{xx}(0)t}),
\end{eqnarray}

from which the frequency of repeating a definite height $(h(x,t)=\alpha)$
can be calculated as

\begin{eqnarray}
\nu_{\alpha}^{+}&=&\int_{0}^{\infty}u P(\alpha,u)du \cr  \nonumber \\
&=&\frac{1}{2\pi}\sqrt{-\frac{D_{xx}(0)}{D(0)}}\exp(-\frac{\alpha^2}{4D(0)t})
\cr  \nonumber \\
&=& \frac{1}{2\pi \sigma} \exp(-\frac{\alpha^2}{4 D_0 D(0)t})
\end{eqnarray}

The quantity $\nu _\alpha ^{+}$ in RD model has a gaussian form with respect to $
\alpha $. The zero level crossing scales with $\sigma $ as $\nu _{\alpha
=0}^{+}\sim \sigma ^{-1}$. Also using the eq.(22) it is found that $
N_{tot}^{+}=D_0^{1/2}\pi ^{-3/4}\sigma ^{-3/2}t^{1/2}$. This shows that
there is no stationary state for the RD model and the quantity $
N_{tot}^{+}$ diverges without saturation.

\section{Frequency of a Definite Height With Positive Slope for KPZ Equation
Before the Singularity Formation}

In the Kardar-Parisi-Zhang (KPZ) model (e.g. in the 1-dimension), the
surface height $h(x,t)$ on the top of location $x$ of 1-dimensional
substrate satisfies a stochastic random equation:

\begin{equation}
\frac{\partial h}{\partial t} - \frac{\alpha}{2}(\partial_x h)^2 = \nu
\partial^ 2_x h+f(x,t).
\end{equation}

where $\alpha \geq 0 $ and $f$ is a zero-mean, statistically homogeneous,
white in time and gaussian process with covariance as  eq.(16).  The
parameters $\nu$, $\alpha$ and $D_0$ ( and $\sigma$) are describing surface
relaxation, lateral growth and the noise strength, respectively. Let us
define the generating function $Z(\lambda, \mu,x,t)$ as:

\begin{center}
$ Z(\lambda,\mu,x,t) = \langle \exp(-i\lambda (h(x,t)-\bar h)-i\mu
u(x,t))\rangle $.
\end{center}

Where $u(x,t)=-\partial_x h(x,t)$. Assuming statistical homogeneity i.e. $
Z_x = 0$ it follows from eq.(23) that $Z$ satisfies the following equation;

\begin{eqnarray}
&-&i\mu Z_t=\gamma(t)\lambda\mu Z-\frac{\alpha}{2}\lambda\mu Z_{\mu\mu}
+i\lambda^2\mu k(0)Z \cr  \nonumber \\
&-&i\mu^3k_{xx}(0)Z -i(\nu\lambda^2+i\alpha\lambda)Z_{\mu} \cr  \nonumber \\
&-&\mu^2\nu\langle u_{xx}(x,t)\exp(-i\lambda \tilde h(x,t)-i\mu
u(x,t))\rangle.
\end{eqnarray}

where $k(x-x^{\prime}) = 2 D_0 D(x-x^{\prime}) $, $\gamma(t)={\bar h}_t$, $
k(0) = \frac{D_0}{\sqrt{\pi} \sigma}$ and $k_{xx}(0) = -\frac{2D_0}{\sqrt{\pi
} \sigma^3}$. and ${\tilde h(x,t)}=h(x,t)-\bar h$. Once trying to develop
the statistical theory of the roughened surface it becomes clear that the
inter-dependency of the height difference and height gradient statistics
would be taken into account. The very existence of the non-linear term in
the KPZ equation leads to development of the cusp singularities in a {\it
finite time} and in the strong coupling limit i.e. $\nu \rightarrow 0$. So
one would distinguish between different time regimes. Recently it has been
shown that starting from the flat interface the KPZ equation will develop
sharp valleys singularity after time scale $t*$, where $t*$ depends on the
forcing properties as $t* = (\frac{1}{4})^{4/3} (\pi)^{1/6} {D_0}^{-1/3}
\alpha^{-2/3} \sigma^{5/3} $ [10]. This means that for time scales before $t*
$ the relaxation contribution tends to zero when $\nu \rightarrow 0$. In
this regime one can observe that the generating function equation is closed.
The solutions of the resulting equation is easily derived, ( starting from a
flat surface, i.e. $h(x,0)=0$ and $u(x,0)=0$), and has the following form
[10]; 
\begin{eqnarray}
&&Z(\mu, \lambda , {t})= ( 1 - {\tanh}^{2}(\sqrt{2i{{k}_{{xx}}}(0)\alpha
\lambda}{t})) \cr  \nonumber \\
&&\exp[-\frac{5}{8}\ln(1-\tanh^{4}(\sqrt{2i{{k}_{{xx}}}(0)\alpha \lambda}{t}
)) \cr  \nonumber \\
&+&\frac{5}{4}\tanh^{-1}(\tanh^{2}(\sqrt{2ik_{xx}(0)\alpha\lambda}
t))-\lambda^{2}k(0) t \cr  \nonumber \\
&-&\frac{1}{16}\ln^{2}(\frac{1-\tanh(\sqrt{2ik_{xx}(0)\alpha\lambda}t)} {
1+\tanh(\sqrt{2ik_{xx}(0)\alpha\lambda}t)}) \cr  \nonumber \\
&-&\frac{1}{2}i\mu^{2}\sqrt{\frac{2ik_{xx}(0)}{\alpha\lambda}}{\tanh}(\sqrt{
2i{{k}_{{xx}}}(0)\alpha \lambda}{t})].
\end{eqnarray}
One can construct $P({\tilde h},u,t)$ in terms of generating function $Z$ as
eq.(19), from which the frequency of repeating a definite height $(h(x,t)- 
\bar h =\alpha)$ with positive slope can be calculated as $
\nu_{\alpha}^{+}=\int_{0}^{\infty}u P(\alpha,u)du$.

In fig.2 we plot the $\nu_{\alpha}^{+}$ for time scales before creation of
singularity, $t/t* = 0.05, 0.15$ and $0.25$. In the KPZ equation due to the
nonlinear term there is no $h \rightarrow -h$ symmetry and one can deduce
that the $\nu_{\alpha}^{+}$ is also not symmetric under $h \rightarrow -h$.

To derive the $N_{tot}^{+}$ let us express the $\nu _\alpha ^{+}$ and $
N_{tot}^{+}$ in terms of the generating function Z. It can be easily shown that
the $\nu _\alpha ^{+}$ and $N_{tot}^{+}$ can be written in terms of the
generating function $Z$ as: 
\begin{equation}
\nu _\alpha ^{+}=\frac 1{2\pi }\int_{-\infty }^{+\infty }\int_{-\infty
}^{+\infty }-\frac 1{\mu ^2}Z(\lambda ,\mu )\exp (i\lambda \alpha )d\lambda
d\mu 
\end{equation}
and 
\begin{equation}
N_{tot}^{+}=\int_{-\infty }^{+\infty }-\frac 1{\mu ^2}Z(\lambda \rightarrow
0,\mu )d\lambda .
\end{equation}

Using the eq.( 25 ) one finds $N_{tot}^{+}\sim \sigma
^{-3/2}t^{1/2}$.  We note that the expression of the $\nu _\alpha ^{+}$ for
the RD model and KPZ equation before $t*$ are different functions of $\alpha$ but $
N_{tot}^{+}$ scales as $\sim \sigma ^{-3/2}t^{1/2}$ in both models.
In fig.(3) using the direct numerical integration of 
joint PDF of height and its gradient 
we plot the $N_{tot}$ vs $t$. In this gragh $N_{tot} \sim t^{1/2}$ which is in agreement 
with analytical prediction.

\begin{figure}
\epsfxsize=8.7truecm \epsfbox{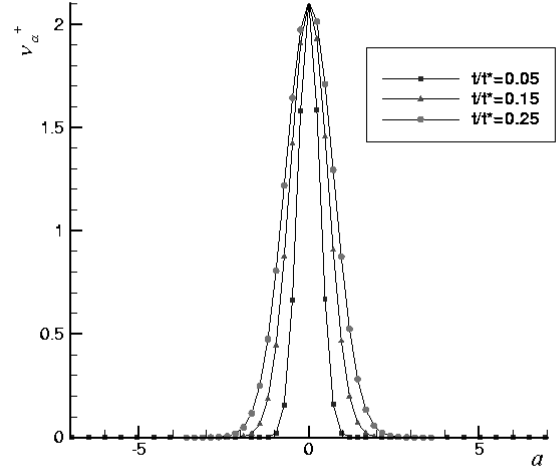} \narrowtext \caption{ plot
of $\nu_{\alpha}^{+}$ vs $\alpha$ for the KPZ equation in the strong coupling 
and befor the creation of sharp vallyes for time scale $t/t* = 0.05, 0.15$ and $0.25$.
}
\end{figure}
 
\begin{figure}
\epsfxsize=8.7truecm \epsfbox{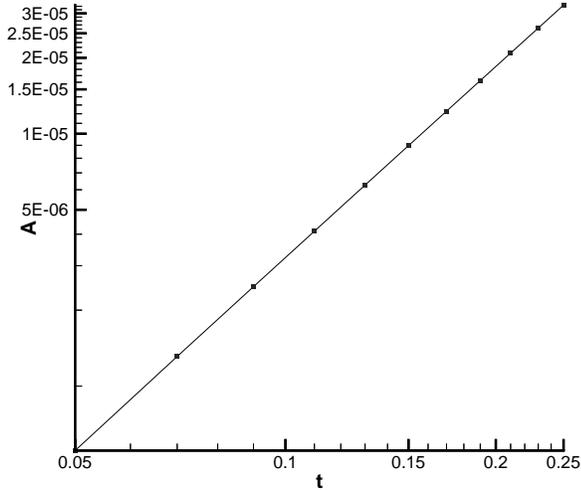} \narrowtext \caption{ log-log plot
of $N_{tot}$ (A) vs $t$ for the KPZ equation in the strong coupling 
 before the creation of sharp vallyes.
}
\end{figure}

\section{conclusion}

 We obtained some results in the problems of RD model and KPZ
equation in 1$+$1 dimensions with a Gaussian forcing which is white in time
and short range correlated in space. We determined the explicit expression of
average frequency of crossing i.e. $\nu _\alpha ^{+}$ of observing of the
definite value for height function $h-\bar{h}=\alpha $ in a growing thin
films for the RD model, from which one can find the averaged number of crossing
the given height in a sample with size L. It is shown the $\nu _\alpha ^{+}$
is symmetric under $h\rightarrow -h$. The integral representation of $
\nu _\alpha ^{+}$ is given for the KPZ equation in the strong coupling limit
before the creation of sharp valleys. We introduced the quantity $
N_{tot}^{+}=\int_{-\infty }^{+\infty }\nu _\alpha ^{+}d\alpha $, which
measures the total number of positive crossing of growing surface and show
that for the RD-model and the KPZ equation in the strong coupling limit and before
the creation of sharp valleys $N_{tot}^{+}$ scales as $\sigma ^{-3/2}t^{1/2}$.
It is noted that for these models $\nu _\alpha ^{+}$ has 
different expression in terms of $\alpha $. The ideas presented in this paper 
can be used to find
the $\nu _\alpha ^{+}$ of the general Langevin equation with given drift and
diffusion coefficients.

{\bf Acknowledgment}
We would like to thank F. Aazami for useful comments.
This work supported by Research Institute for Fundamental Sciences.


\begin{references}
\bibitem{1}  A.-L. Barabasi and H. E. Stanley, ''Fractal Concepts in Surface
Growth'' (Cambridge University Press, New York, 1995).

\bibitem{2}  T. Halpin-Healy and Y. C. Zhang, Phys. Rep.{\bf 245}%
,218(1995);J. Krug, Adv. Phys.{\bf 46},139(1997)

\bibitem{3}  J. Krug and H. Spohn in ''Solids Far from Equilibrium Growth,
Morphology and Defects'', edited by C. Godreche (Cambridge University Press,
New York, 1990).

\bibitem{4}  P. Meakin, ''Fractal, Scaling and Growth Far from Equilibrium''
(Cambridge University Press, Cambridge, 1998).

\bibitem{5}  M. Marsilli, A. Maritan, F. Toigoend J.R. Banavar,Rev. Mod.
Phys., {\bf 68},963 (1996)

\bibitem{6}  M. Kardar, Physica A {\bf 281},295(2000).

\bibitem{7}  F. Family and T. Vicsek,J.Phys.A {\bf 18},L75(1985)

\bibitem{8}  S.O. Rice '' Mathematical Analysis of Random Noise'' Bell
System Tech. J. Vol.23,(1944),282; Vol. 24, (1945), 46

\bibitem{9}  D.E. Newland '' An Introduction to Random Vibration, Spectral
and Wavelet Analysis'' ( Longman, Harlow and Wiley, New York, 1993 )

\bibitem{10}  A. A. Masoudi, F. Shahbazi, J. Davoudi and M. Reza Rahimi
Tabar, Phys. Rev. E {\bf 65}, 026132(2002).
\end{references}
\end{document}